\documentclass[twocolumn,showpacs,preprintnumbers,amsmath,amssymb]{revtex4}
\usepackage{tabularx,graphicx}\begin{document}
\newcommand{\beq}{\begin{equation}}
\newcommand{\eeq}{\end{equation}}
\newcommand{\beqn}{\begin{eqnarray}}
\newcommand{\eeqn}{\end{eqnarray}}
\newcommand{\bmath}{\begin{subequations}}
\newcommand{\emath}{\end{subequations}}
\title{Hole superconductivity in $H_2S$ and other sulfides under high pressure}
\author{J. E. Hirsch$^{a}$ and F. Marsiglio$^{b}$ }
\address{$^{a}$Department of Physics, University of California, San Diego,
La Jolla, CA 92093-0319\\
$^{b}$Department of Physics, University of Alberta, Edmonton,
Alberta, Canada T6G 2J}
\begin{abstract}
Superconductivity at temperatures up to 190 K at high pressures has recently been observed in $H_2S$ and interpreted as conventional BCS-electron-phonon-driven superconductivity.\cite{h2s} Instead we propose  that it is another example of the mechanism of hole superconductivity at work. Within this mechanism high temperature superconductivity arises when holes conduct through negatively charged anions in close proximity. We propose that electron transfer from $H$  to $S$ leads to conduction by
holes in a nearly full band arising from direct overlap of $S^=$ $p$ orbitals in a planar structure. The superconductivity is non-phononic and is driven by
 pairing   of heavily dressed hole carriers to lower their kinetic energy. Possible explanations for the observed lower critical
 temperature of $D_2S$  are discussed. We predict that 
 high temperature superconductivity will also be found in other sulfides under high pressure such as 
 $Li_2S$, $Na_2S$ and $K_2S$. \end{abstract}
\maketitle

\section{Introduction}
The origin of superconductivity in many materials  remains controversial. However, there exists near universal consensus that for a large class of materials, termed `conventional superconductors',   the electron-phonon 
interaction is  the driving force for superconductivity.\cite{bcs} These materials are described by BCS-Eliashberg theory.\cite{eph,parks} When a new superconductor
 is found, one of the first questions asked
is whether it is `conventional' or `unconventional'. In recent years, a growing number of superconductors have been classified as `unconventional' 
because their properties differ from those expected from a  BCS electron-phonon superconductor.\cite{norman} Often, the critical temperature is higher 
than can be accounted for by the electron-phonon mechanism.

Instead, within the theory of hole superconductivity \cite{holews}  superconductivity is driven by a universal mechanism that is  $not$ the electron-phonon
interaction.\cite{holesc} When a new superconductor is found, the first  question to be asked is, does this material prove  the theory wrong?  The theory of hole superconductivity is falsifiable
by finding a single material that clearly does not conform to it.

Recently,\cite{h2s} superconductivity at high pressures at temperatures up to $T_c=190K$, higher than for any previously known superconductor,
 was reportedly observed in $H_2S$, with somewhat lower temperatures for $D_2S$, and it was proposed that it is conventional superconductivity driven by the electron-phonon interaction, with the high
$T_c$ arising because of the light mass of the $H$ atom.\cite{h2s} In this paper we propose instead that the observed
superconductivity is described by the theory of hole superconductivity.

The theory of hole superconductivity predicts that high temperature superconductivity results from  a few $holes$ 
conducting through a network of closely spaced negatively charged $anions$, in conducting substructures with excess negative charge.\cite{matmec}
Within this theory, superconductivity in the high $T_c$ cuprates arises from holes conducting through direct hopping between
$O^=$ anions in the  negatively charged $CuO_2$ planes,\cite{cupr,cupr5} with the $Cu^{++}$ cations playing a secondary role; superconductivity in $MgB_2$ arises from holes conducting through direct hopping between
$B^-$ anions,\cite{mgb2}  with the $Mg^{++}$ cations playing a secondary role; and superconductivity in the iron pnictide and iron chalcogenide materials
arises from holes conducting through direct hopping between
$As^{---}$, $S^=$ or $Se^=$ anions\cite{iron,matmec}, with the $Fe^{++}$ cations playing a secondary role. Similarly, we propose here  that 
superconductivity in $H_2S$ under pressure arises from holes conducting through direct hopping between
$S^=$ anions, with the $H^+$ cations playing a secondary role.

 Metallization of $H_2S$ under $96GPa$ pressure was reported in Ref. \cite{saka} in 1997. It was proposed as a likely
 mechanism that the molecules dissociate at this pressure and an atomic arrangement results, bringing $S$ atoms in contact
 with each other and the hydrogen occupying interstitial sites. (Figure 4 (a) in Ref. \cite{saka}). In Ref. \cite{li}, a density functional calculation predicted 
 a variety of different phases as a function of pressure. In particular it was found that at pressure $160GPa$ 
 a transition from a molecular to an atomic arrangement would occur. The atomic Cmca structure found (Fig. 1(e)  of ref. \cite{saka}) clearly allows for conduction through direct hopping between $S^=$ ions in the $a$ and $b$ directions, while
 conduction in the $c$ direction would be through intermediate $H^+$ ions. Thus our theory predicts that conduction in the
 $a-b$ `planes' in this structure would drive superconductivity.
 
 Superconductivity is reported to appear shortly after metallization, at pressures of about $100GPa$\cite{h2s}. The superconducting
 transition temperature increases monotonically from about $20K$ to about $150K$ as the pressure is increased to
 $200GPa$. The sample is warmed to $100-150K$ before the pressure is increased to the next value, then it is cooled.
 In addition, another route to superconductivity was found where pressure is applied at significantly
 higher temperatures, $220-300K$. With this procedure, superconductivity is found in the pressure range $P=150GPa$ 
 to $200GPa$ with critical temperature around
 $190K$ weakly dependent on pressure. It is hypothesized that this route results in a different structure of the material\cite{h2s}.

\section{structure}
\begin{figure}
\resizebox{8.5cm}{!}{\includegraphics[width=7cm]{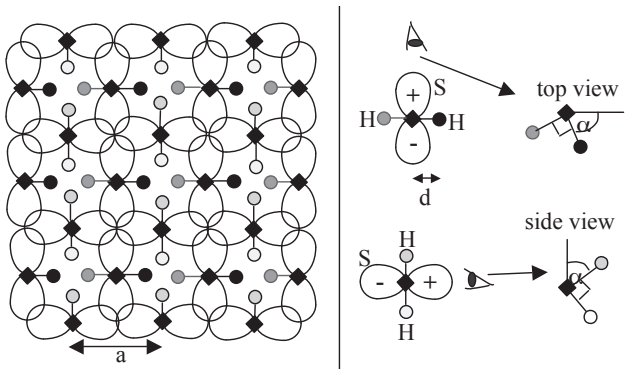}}
\caption {Left panel: proposed lattice structure for superconducting $H_2S$. Planar $p$ orbitals of
the $S^=$ anions (rhombi) overlap allowing for conduction of holes in the plane. 
The hydrogens bonded to the sulphurs are shown as circles with different shadings of grey
indicating their distance to the plane of the paper. The two darker shadings indicate positions
in front of the paper, the two lighter ones positions behind the paper.  The two molecular bonds
to a given $S$ are at $90^o$  to each other and to the corresponding $p$ orbital in the
plane, and the angle between the direction of the bond to the darker circle and the plane of
the paper is denoted by $\alpha$, as indicated in the right panel of the figure.  }
\label{figure5}
\end{figure}

Let us consider how superconductivity may arise within the mechanism of hole superconductivity. Note that
in the $H_2S$ molecule the angle between the two bonds to the $H$ atoms is very close to $90^o$
($92.3^o$). So there are two almost  orthogonal $p$ orbitals involved in the molecular bond,
and the third $p$ orbital is orthogonal to them.
If $S$ atoms arrange themselves in a plane so that the lone $p$ orbital is in the plane, conduction
in the plane can occur through direct hopping between $S$ atoms without involving the $H$.
The structure is shown schematically in Fig. 1. 

Assuming that significant but not complete electron transfer from $H$ to $S$ occurs, the result would be to have highly negatively charged planes of $S^=$ anions where holes will conduct through direct overlap of planar $S$ $p$ orbitals. 
This is precisely the situation that we have proposed exists in the cuprate superconductors and leads to high temperature superconductivity. Increasing the applied pressure will reduce the distance between $S^=$ anions in the plane leading to a larger hopping amplitude $t$ as well
as a larger correlated hopping term $\Delta t$ which would give rise to a higher $T_c$ just as is observed in the cuprates under application of pressure.

\begin{figure}
\resizebox{8.5cm}{!}{\includegraphics[width=7cm]{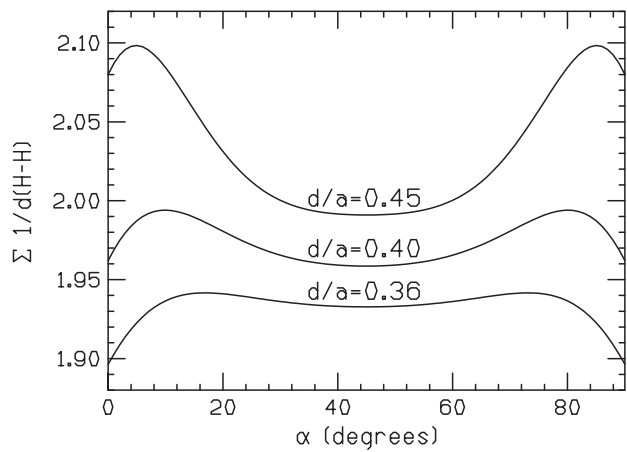}}
\caption { Coulomb interaction between hydrogen ions above and below the plane (arbitrary units). For 
$d/a<0.40$, the lowest energy state corresponds to two hydrogen ions in the plane and two hydrogen ions right above and
below a sulphur ion. For $d/a>0.40$, all the $S-H$ bonds are in directions forming an angle of $45^o$ with the plane. }
\label{figure5}
\end{figure}

The position of the $H^+$ ions relative to the $S^=$ planes may change for different values of
the pressure. We have calculated the Coulomb repulsion between the $H^+$ ions above and
below a plaquette in the plane as a function of the angle $\alpha$ that a bond makes with
the plane as defined in Fig. 1 right panel. Figure 2 shows the result for different values of
$d/a$, with $d$ the $S-H$ bond distance and $a$ the distance between nearest neighbor
$S^=$ ions in the plane. It can be seen that the minimum is at either $\alpha=0$ or
$\alpha=45^o$. As the applied pressure increases the value of $a$ is expected to become
smaller relative to $d$, hence for higher values of the pressure all the $S-H$ bonds
are expected to be at an angle $45^o$ relative to the plane. Instead, for lower pressure
we expect two of the molecular bonds to be in the plane and two perpendicular to the
plane ($\alpha=0$). 
The former situation results in more negative charge in the plane,
which should lead to higher superconducting transition temperatures within the theory of 
hole superconductivity. 

We suggest that a possible interpretation of the two different phases
seen in experiment depending on the annealing route discussed above may correspond to 
these two phases:  when the pressure is applied at lower temperatures the system may adopt the $\alpha=0$ phase for low pressures
since $d/a<0.40$, and may remain in this phase that becomes metastable at higher pressures when $d/a>0.40$. Instead, if the
system is reheated to higher temperatures when the pressure is high it can find its way to the lower energy state $\alpha=45^o$
giving rise to superconductivity at higher temperature.

 \section{atomic parameters and hamiltonian}
 The first and second electron affinities of sulphur are
 \bmath
 \beq
 A(1)=E(S)-E(S^-)=2.08 eV
 \eeq
 \beq
 A(2)=E(S^-)-E(S^=)=-5.54eV
 \eeq
 \emath
 so that the ``effective $U$'' for two holes in the $S^=$ anion is
 \beq
 U_S=E(S^=)+E(S)-2E(S^-)=7.62 eV
 \eeq
 This value is smaller than the effective $U$ for two holes in the $O^=$ anion. In that case,
 $A(1)=1.46eV$, $A(2)=-8.79eV$ and the effective $U$ is
 \beq
 U_O=E(O^=)+E(O)-2E(O^-)=10.25 eV
 \eeq
 In the theory of hole superconductivity the pair wave function always has some amplitude for on-site pairing and as a consequence
 the on-site $U$ suppresses superconductivity. These values of $U$ for oxygen and sulphur suggest that 
 sulphur compounds can potentially exhibit higher $T_c's$ than the high $T_c$ cuprates due to the smaller value of $U_S$
 versus $U_O$.\cite{hole1}
 
 Note also that the radius of the $S^=$ anion is $1.84 \AA$, versus $1.40 \AA$ for the $O^=$ anion. This suggests that it should be easier to attain
 structures under pressure where neighboring  anion orbitals overlap for $H_2S$ than it would be for $H_2O$, thus favoring  metallicity and superconductivity
 in the former over the latter occurring through direct hopping of carriers between the anions, as required in the theory under discussion here.
 Note also that because of its  larger size the polarizability of the $S^=$ anion is over twice higher than that of $O^=$ which will
 favor superconductivity by reducing the importance of direct Coulomb repulsion between carriers.

The interaction term leading to pairing in the theory of hole superconductivity is the correlated hopping interaction $\Delta t$,
which in its simplest form is given by
\beq
\Delta t=tS(1-S)
\eeq
where $t$ is the hopping amplitude for one electron in the empty band and $t_h=tS^2$ is the hopping amplitude for one hole in the full
band. $S<1$ is given by the overlap matrix element of the atomic orbital when one electron and two electrons are in the orbital.
The ratio between correlated hopping amplitude and single hole hopping amplitude is
\beq
\frac{\Delta t}{t_h}=\frac{1}{S}-1\equiv \Upsilon
\eeq
which is the coupling strength driving superconductivity, and becomes larger the smaller the overlap matrix element $S$ is.

The atomic p-orbital is given by
\beq
\varphi_\alpha(r)=\sqrt{\frac{8\alpha^5}{3\pi}}r(1-\frac{\alpha r}{2})cos\theta e^{-\alpha r}
\eeq
using atomic units ($a_0=1$). For atomic charge $Z$ and a single electron in the orbital, the energy is
\beq
E(1)=\frac{\alpha^2}{2}-\frac{\alpha Z}{3}
\eeq
where the first term is potential and the second kinetic energy. Minimization with respect to $\alpha$ gives $\alpha=Z/3$ for the
state of minimum energy with one electron in the orbital. When there are two electrons in the orbital the expectation value of
the energy is
\beq
E(2)=\alpha^2-\frac{2\alpha Z}{3}+c\alpha
\eeq
where the last term is the expectation value of the Coulomb repulsion between electrons, with  $c$ a numerical constant. The energy is now minimized
by
\beq
\bar{\alpha}=\frac{Z}{3}-\frac{c}{2}
\eeq
giving the orbital exponent of the expanded orbital with two electrons. If we assume that Slater shielding rules are satisfied \cite{slater},
$c=0.35\times 2/3=0.233$.

The overlap matrix element between the expanded and
unexpanded orbital is then
\beq
S=\frac{  (\alpha \bar{\alpha})^{5/2}       (3\alpha \bar{\alpha}-\alpha^2-\bar{\alpha}^2)    }       {(\frac{\alpha+\bar{\alpha}}{2})^7}
\eeq
and it approaches zero when $Z\rightarrow Z_c=3c/2=0.35$.
In the case considered here, the effective value of $Z$ depends on the amount of negative charge transfer from the two $H$ atoms
to the $S$ atom bound to them. For complete charge transfer, $Z$ approaches zero. Assuming significant charge transfer
$Z$ and as a consequence $S$ become  small, giving rise to a large coupling parameter $\Upsilon$.

\section{hamiltonian and critical temperature}
The effective Hamiltonian for hole conduction through $S^=$ anions is given by \cite{cupr}
\bmath
\beq
H=-\sum_{ij\sigma}  t_{ij}^\sigma [ {c}_{i\sigma}^\dagger  {c}_{j\sigma}+h.c.]+U \sum_i  {n}_{i\uparrow} {n}_{i\downarrow}
\eeq
\beq
t_{ij}^\sigma=t_h+\Delta t(n_{i,-\sigma}+n_{j,-\sigma})
\eeq
\emath
and it gives rise to superconductivity provided the condition
\beq
S\leq \sqrt{1-U/(2zt)}
\eeq
is satisfied,\cite{dynh,cupr} with $z=4$ the number of nearest neighbors. For example, for bare hopping $t=1eV$ Eq. (12) yields $S<0.22$ for 
superconductivity to occur for $U=U_S$ [Eq. (2)]. The critical temperature increases as the inequality in Eq. (12) grows.

In particular, application of pressure increases the value of $t$ by reducing the interatomic distances and this makes the condition
Eq. (12) less stringent.  Pressure is not expected to significantly change the values of $S$ [Eq. (10)] or $U$ which are atomic properties, insofar as the 
charge transfer between $H$ and $S$ atoms is not significantly affected. Thus, pressure will increase the critical temperature for given
band filling assuming no change in the lattice structure occurs.

\begin{figure}
\resizebox{8.5cm}{!}{\includegraphics[angle=-90,width=7cm]{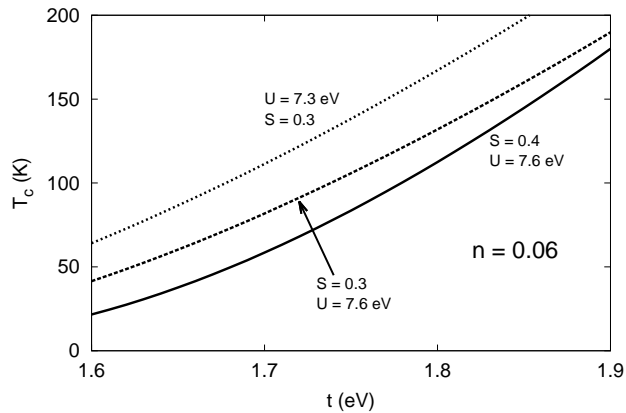}}
\caption { $T_c$ vs hopping amplitude $t$ for three different sets of parameters, $S=0.4$ and $U = 7.6$ eV (solid curve),
$S=0.3$ and $U=7.6$ eV (dashed curve), and $S=0.3$ and $U=7.3$ eV.}
\label{figure3}
\end{figure}

The behavior of the critical temperature versus hopping amplitude $t$ is shown in Figure 3 for various values of the parameter $S$,
for hole doping $n_h=0.06$, which is close to optimal hole doping for this model.\cite{cupr}. For other band fillings the
behavior is  similar. Here we have used $S=0.4$ and $U = U_S \approx 7.6$ eV to illustrate the typical trend with pressure (solid curve). 
$T_c$ always rises with pressure, assuming no effects beyond increasing $t$ as discussed above. Reducing the value of $S$ tends to
increase $T_c$, and a similar curve is illustrated for the same Coulomb repulsion, but $S=0.3$ (dashed curve). 
Finally, reducing the Coulomb repulsion raises $T_c$, as indicated by the dotted curve; in all three cases the behaviour with pressure is
very similar. Not shown is the dependence on hole density; this is more complicated and our theory predicts an optimal value as a
function of $n_h$, \cite{cupr} but for the sulfides there may be additional complexities.

\section{isotope effect}

The measured isotope effect has been cited as evidence for electron-phonon driven superconductivity.\cite{h2s} However,
the observed decrease in superconducting $T_c$ is far too large to be explained by the electron-phonon mechanism. Using
$T_c \approx M^{-\alpha}$ to define the isotope coefficient, $\alpha$, so that the `BCS' value is $\alpha = 0.5$, we note that
the observed $\alpha \approx 1.0$ is significantly larger. Can this be understood within the conventional Eliashberg framework?
The answer is a definitive `No!' Most `conventional' superconductors have an isotope coefficient somewhat lower than $0.5$, and
there are various reasons for why this might be consistent with electron-phonon driven superconductivity. Garland \cite{garland63}
first noted that the presence of a direct Coulomb repulsion between two electrons will generally reduce the isotope coefficient from $\alpha = 0.5$
and values approaching $\alpha \approx 0$ are possible, provided $T_c$ is quite low on the scale of the Debye frequency. The physics is that
it is not so much the attraction due to the phonon-mediated interaction between two electrons that is important; rather it is the retardation of this
induced interaction compared to the direct Coulomb repulsion that is key for pairing. As this begins to play more of a role the change in ionic mass
has less impact on $T_c$ and the isotope effect diminishes.

Other factors tend to decrease the isotope coefficient as well. For example, a ``differential isotope exponent'' \cite{rainer79} discriminates 
on the basis of phonon frequency the relative contributions to the total isotope effect. In cases where a large mass discrepancy 
exists in the constituents of a superconductor consisting of several types of atoms, then high frequency components in the electron-phonon coupling
spectrum will be strongly associated with vibrations involving the lighter mass. This is the case in palladium-hydride\cite{stritzker72}
where H is much lighter than Pd, and is also the case here. Rainer and Culetto \cite{rainer79} (see also \cite{ashauer87,knigavko01} for an
analysis of the cuprates and MgB$_2$, respectively) found that substitution of selected elements tends to result in a reduced isotope effect, but
the sum of these results in a total isotope coefficient, $\alpha_{\rm tot} = 0.5$ (in the absence of direct Coulomb repulsion).  

It should be remembered that even in the relatively simple case of superconducting elements the isotope coefficient is sometimes 
anomalous \cite{allen} and for some cases has not been convincingly explained:  ruthenium \cite{ru} and zirconium \cite{zr} show an isotope coefficient 
$\alpha\sim 0$, and for $\alpha-$uranium the isotope exponent is found to be $\alpha=- 2$. \cite{ur}

Indeed, the case of Pd-H/D \cite{stritzker72} is similar to the present case (H$_2$S) as far as the isotope effect is concerned, so it is noteworthy that a
large {\it negative} isotope coefficient was observed in that case. \cite{stritzker72} Most workers in the field have reconciled this anomaly
with the presence of large anharmonic effects, and the increase in coupling strength that is caused by these effects when Deuterium is substituted
for Hydrogen.\cite{ganguly75,klein77}. More recent calculations that confirm this picture can be found in Refs. \cite{klein92,errea13}.

Anharmonicity is expected to play an even greater role in H$_2$S, because of the much higher temperatures involved. Yet, the measured
isotope coefficient, while anomalous, is anomalous in the exact opposite direction. This is a less common occurrence in superconducting
materials. The most prominent  such example is the case of the underdoped cuprates, where a very large isotope effect ($\alpha {{ \atop >} \atop 
{ \sim \atop }} 1.0$)
is observed.\cite{franck94} In the case of the cuprates, nobody really believes that the large isotope effect is connected to the electron-phonon
interaction.

This brings us to H$_2$S, and the anomalously high isotope coefficient. It is unlikely due to anharmonicity, as the previous work summarized
above indicates that anharmonicity should lead to an {\it enhancement} of $T_c$ for the heavier isotope. 
Thus, we argue that the observed behavior does $not$  support the scenario where the high $T_c$ in this material
is attributable to the electron-phonon interaction and the light mass of $H$.\cite{h2s}
 
 For the mechanism  of hole superconductivity, a positive isotope effect is generically expected because an increased amplitude of zero-point
 ionic vibrations   increases the mean square hopping amplitude and correspondingly also the correlated hopping interaction 
 $\Delta t$. We have proposed that this is the origin of the positive isotope effect upon $B$ substitution seen in $MgB_2$.\cite{mgb2}
 However, for the present case we do not expect the dominant conduction channel giving rise to high $T_c$ to involve the
 cations $H$ or $D$, so that this argument would not apply.
 
 Instead we conjecture as a possible explanation for the observations that for the case of the lighter cation ($H$) an increased charge
 electron transfer from the cation to the sulphur occurs due to the larger amplitude of the $H$ zero point motion that gets it into closer
 proximity to the $S$. Such an increased charge transfer would both reduce the effective Coulomb repulsion on the sulphur anions
 and decrease the value of the overlap matrix element Eq. (10), with both effects contributing to a higher $T_c$ according to
 Eq. (12). 
 
 Alternatively it is possible that the experimental results reported so far are misleading and in fact the $T_c$'s reported for 
 $D_2S$ and $H_2S$ correspond to different structures of the material under pressure and the difference in $T_c$ originates
 principally in the different structures rather than any difference in the vibrational frequencies of the cations. We believe it is likely that a 
 higher $T_c$ phase in the $D_2S$ material under pressure will be found in future experiments.

\section{discussion}

Within the theory of hole superconductivity, lattice stability and superconductivity compete with each other \cite{matthias} because
{\it antibonding electrons} tend to break the lattice apart.\cite{hole2} Antibonding states are those near the top of electronic energy bands, which are occupied
in the regime where superconductivity is favored within this theory. Application of external pressure can stabilize lattice structures with
band fillings favorable to hole superconductivity (i.e. nearly filled bands) by counterbalancing the pressure exerted by the antibonding
electrons.

It was clear from its inception \cite{hole1} that within this theory superconductivity is favored when conduction occurs through atoms
on the right side of the periodic table, which normally form insulators rather than metals, and in particular when conduction occurs
through holes in negatively charged anions with filled shells. In our initial paper \cite{hole1} we stated: ``the fact that the free ion on-site repulsion for  
two holes on $S^{2-}$ is only 7.6 eV and the polarizability  of $S^{2-}$ is over twice higher than that of $O^{2-}$ suggests that very high $T_c$ superconductivity is possible in a  
hole conductor with $S^{2-}$ anions'', foreshadowing the recent discovery.\cite{h2s} The large positive pressure dependence of $T_c$ observed
in the high $T_c$ cuprates is naturally explained in our theory arising from the increased overlap between neighboring anion orbitals caused
by their close proximity,\cite{cupr}
which suggests quite generally that application of high pressure is a route to high temperature superconductivity within this theory.

Putting all these considerations together it is clear that within the theory of hole superconductivity natural candidates for high $T_c$ are
insulating materials containing atoms from the right side of the periodic table (columns V, VI and VII) rendered superconducting by application of
high pressure. Binary compounds with polar covalent bonds or ionic bonds are more favorable than single elements because they allow for the possibility of having {\it negatively charged anions}, whose orbitals will overlap under application of sufficiently high pressure.   
In particular, insulating compounds that would  be candidates for high temperature superconductivity under high pressure within this theory
are $Li_2S$,  $Na_2S$ and  $K_2S$. Under even higher pressure, high temperature superconductivity
may be achieved 
 in  $H_2O$, $Li_2O$,  $Na_2O$ and  $K_2O$. The absence of a clear downtrend in $T_c$ as the mass of the cation increases, which would mirror
the absence of such downtrend in the superconducting transition temperature of elements in the periodic table,\cite{buzea} would establish
that the high temperature superconductivity of $H_2S$ does $not$ originate in the light mass of $H$ \cite{h2s,ashcroft} but rather that the
key component is  the $S^=$ anion.

\begin{acknowledgments}

FM is grateful for support from
the Natural Sciences and Engineering Research Council of Canada (NSERC),
and by the Alberta iCiNano program.

\end{acknowledgments}

 \end{document}